\documentclass[9pt,journal,compsoc]{IEEEtran}
\usepackage{graphicx}
\usepackage{balance}
\ifCLASSOPTIONcompsoc
  \usepackage[nocompress]{cite}
\else
  \usepackage{cite}
\fi
\ifCLASSINFOpdf
\else
 \fi
\begin{document}

\title{\huge The Case for a Wholistic Serverless Programming Paradigm and Full Stack Automation for AI and Beyond -- The Philosophy of Jaseci and Jac}

\author{
    \IEEEauthorblockN{Jason Mars\IEEEauthorrefmark{1}\IEEEauthorrefmark{2}}
    \\
    \IEEEauthorblockA{\IEEEauthorrefmark{1}University of Michigan, \emph{profmars@umich.edu}}
    \\
    \IEEEauthorblockA{\IEEEauthorrefmark{2}Jaseci Labs, \emph{jason@jaseci.org}}

}

\markboth{Position Paper -- June~2022}%
{Shell \MakeLowercase{\textit{et al.}}: Bare Demo of IEEEtran.cls for Computer Society Journals}

\IEEEtitleabstractindextext{%
    \begin{abstract}

        Modern production applications are \emph{diffuse}, spanning multiple individual programs (database, memcache, logging, application logic, AI models, etc) interfacing each other over APIs to realize a single product functionality.
        Creating such applications at scale is technically challenging, requires a highly-skilled developer team, is rife with complexity, and is, for many, prohibitively costly.
        This complexity is in stark contrast to the era of computing where a state of the art software product was a single binary that ran on one machine and could be developed by a single programmer.
        Though a number of important abstractions and technologies have emerged to help mitigate this complexity, the creation of sophisticated production software in practices is still highly complex and requires a team of engineers.

        In this work, the case is made for a wholistic top-down re-envisioning of the system stack from the programming language level down through the system architecture to bridge this complexity gap.
        The key goal of our design is to address the critical need for the programmer to articulate solutions with higher level abstractions at the problem level while having the runtime system stack subsume and hide a broad scope of diffuse sub-applications and inter-machine resources.
        This work also presents the design of a production-grade realization of such a system stack architecture called \textbf{Jaseci}, and corresponding programming language \textbf{Jac}.
        Jac and Jaseci has been released as open source and has been leveraged by real product teams to accelerate developing and deploying sophisticated AI products and other applications at scale.
        Jac has been utilized in commercial production environments to accelerate AI development timelines by  $\sim$10x, with the Jaseci runtime automating the decisions and optimizations typically falling in the scope of manual engineering roles on a team such as what should and should not be a microservice and changing those dynamically.
    \end{abstract}

    }

\maketitle

\IEEEdisplaynontitleabstractindextext

\IEEEpeerreviewmaketitle

\IEEEraisesectionheading{\section{Introduction}\label{sec:introduction}}

\begin{figure}[tb]
    \centering
    \includegraphics[width=\linewidth]{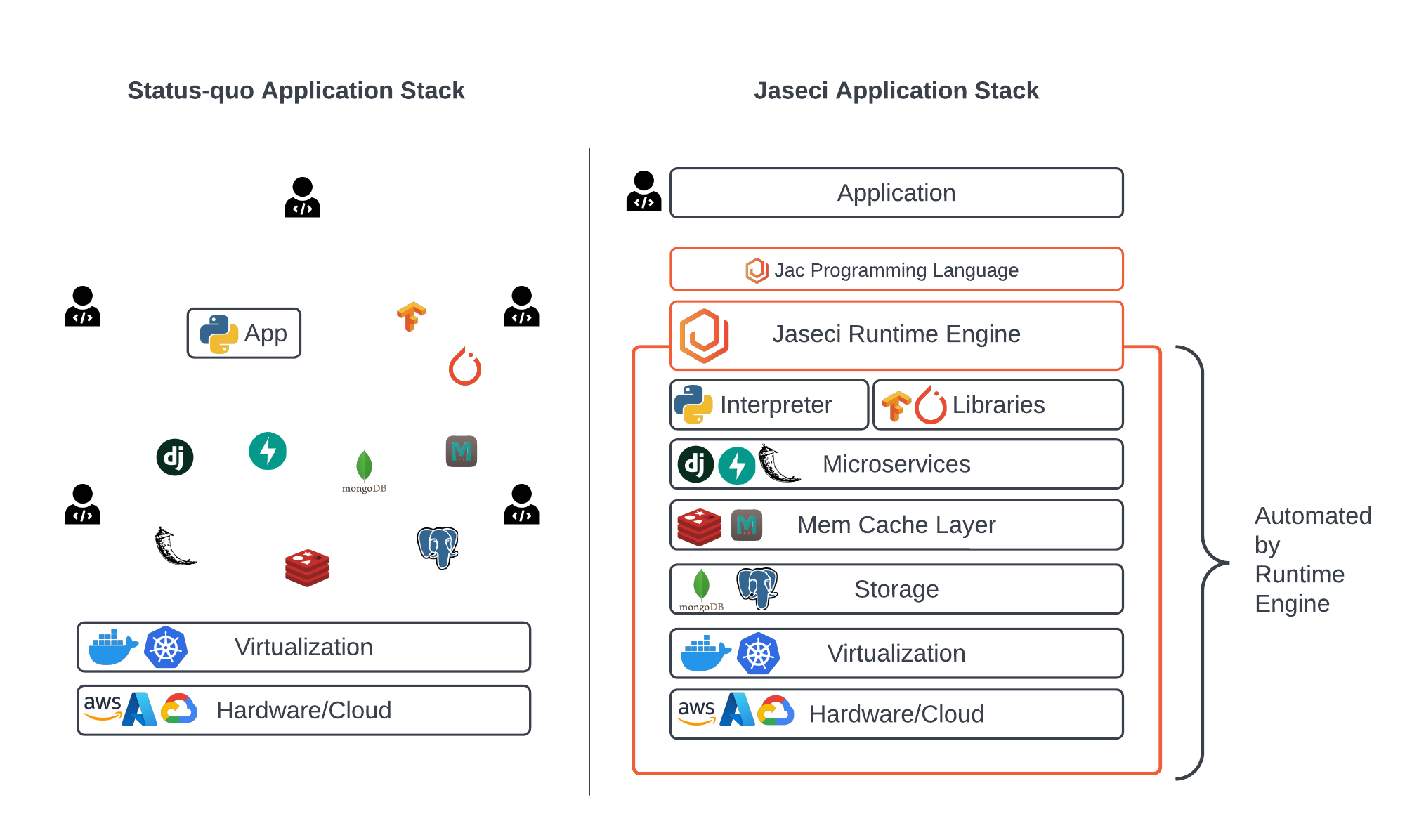}
    \caption{Comparison between status quo development of production grade \emph{diffuse} applications (left), and the Jaseci technology stack that hides and automates an expanded set of subsystems through raising the level of abstraction (right).}
    \label{fig:intro}
\end{figure}

\IEEEPARstart{T}{here} has been a fundamental paradigm shift in the landscape of how we build software over the last 2 decades.
The compute stack was originally envisaged with the assumption that a single program would run on a single machine.
In this traditional model, system software abstractions subsumed the management of processor, memory, disk and physically connected peripherals within the context of the machine.
However, this landscape rapidly changed with the evolution toward software being served on the backbone provided by the internet.
Now, an `application' is realized through the cooperation of multiple distinct sub-applications (services) running collaboratively.
For example a single application my contain self one or more self-contained database, memcache, logging, application logic, AI model applications interfacing each other over APIs as shown in Figure~\ref{fig:intro} (left).
We call these applications \emph{diffuse applications}.

This work contends that the fundamental programming paradigms in computing has not evolved at pace.
The abstractions envisioned during the era of the single machine computational model is still present at the programming interface and throughout the runtime stack leading to significant and costly complexity.

To address this complexity, two keystone abstractions have recently emerged to facilitate the development of these diffuse applications.
The first of these abstractions is the introduction and rapid dissemination of containerization service platforms.
With what started as a key insight articulated in ``The Datacenter as a Computer,'' Google would innovate their Borg system and ultimately released it open source as Kubernetes.
With Kubernetes, the underlying hardware resources would be abstracted away with the introduction of \emph{pods} (virtual machines), and other resources that can be virtually networked together and otherwise configured irrespective of the physical hardware.
Today, Kubernetes is the most prevalent containerized service abstraction layer in cloud computing.
The second of keystone abstraction would be coined ``Severless Computing'' and gained prominence with the introduction of Amazons Lamda functions.
This FaaS abstraction would facilitate the development of diffuse applications at the level of functions and abstract away the underlying containerized service ecosystem.
A programer can simply make function calls in their favorite language without every needing to be aware of where the function will run nor the system level resources that would be allocated or managed.

Though these two abstractions have been highly impactful, these innovations in our stack architecture represent a bottom up evolution of abstractions.
As a result, programmers are still left with single-machine abstractions at the programming interface and must grapple with a significant amount of complexity.
For example, traditional languages and their runtime stacks are predominately designed with the goal of hiding and managing intra-machine resources while what is needed for diffuse applications is the hiding and management of inter-machine resources.
Analogous to the virtualization and management of allocated memory on the heap provided by garbage collectors in modern languages (intra-machine), the virtualization and management of resources such as microservice creation, scheduling and orchestration alongside policies for organizing distributed databases, mem caches, logging and other highly complex subsystems (inter-machine) is not only needed, but as we show in this work, possible and practical.
Without this raising of the level of abstraction, it has become prohibitively difficult for a single engineer to invent, build, deploy, launch, and scale modern cutting edge applications.

To the best of our knowledge, we are not aware of a thorough, wholistic, and top-down design of a serverless programming paradigm and computational stack from the language level down through the system runtime stack to hide this expanded set of resources.

In this work, we present a wholistic design approach with the goal of abstracting away and automating a new class of underlying systems, allowing a programmer to articulate solutions and diffuse applications at the problem level.
We present the design of a \emph{diffuse runtime execution engine} we call \textbf{Jaseci}, and a \emph{data-spacial programming language} we call \textbf{Jac}.
The design of Jaseci and Jac has initially been inspired to by sophisticated emerging AI applications at scale and is driven by two key insightss.
\begin{itemize}

    \item \emph{Higher level abstractions are needed at the language level to allow single creators to work at the problem level to build end-to-end diffuse AI products.}
    \item \emph{A new set of abstractions across the language runtime and system stack is needed to automate and hide the class of inter-machine resources from the programmer.}

\end{itemize}

\noindent
To this end we present techniques across two categories,

\begin{enumerate}

    \item \emph{Jac Language} - A language that introduces a new set of abstractions, namely \textbf{data-spacial scoping} and \textbf{agent oriented programming}. These abstractions natively facilitates the emerging need to reason about and solve problems with graph representations as well as the need for algorithmic modularity and encapsulation to hide a new class of inter-machine resources.
    \item \emph{Jaseci Diffuse Runtime Engine} - A runtime that raises the abstraction layer to the problem solving level where the runtime engine subsumes responsibility for not only for the optimization of program code, but the orchestration, configuration, and optimization of the full cloud compute stack and inter-machine resources (such tasks as container formation, scaling and optimization).

\end{enumerate}

Jaseci and Jac is fully functional, open-source~\cite{jaseci-website,jaseci-github,jaseci-pypi}, and used in production for four real-world products today.
These commercial products were built entirely on the Jaseci staci and includes Myca~\cite{myca-website}, HomeLendingPal~\cite{hlp-website},  ZeroShotBot~\cite{zsb-website} and TrueSelph~\cite{ts-website}.
Across these and other projects, the Jac language has been used by dozens of programmers in the creation of production software and Jaseci deployments support tens of thousands of production queries per day currently.
In practice, our initial infrastructure has been leveraged in practice to achieve 10x reduction in development time and near 100\% elimination of typical backend code needed for a complicated AI based application.

The specific contributions of this paper include:
\begin{itemize}
    \item We formulate the problem of development complexity and present a top down programing paradigm and runtime stack for diffuse applications.
    \item We describe the design and implementation Jaseci's \textbf{diffuse runtime execution engine}.
    \item We introduce Jac, a language that implements a \textbf{data-spacial} programming paradigm (the first of its kind).
    \item We describe the utility of Jaseci and Jac through real world case studies of building out a real production scale-out product.
\end{itemize}

We find that the wholistic design philosophy and resulting paradigm of Jaseci and Jac is a promising one.
Multiple development teams have adopted the data spacial programming model of Jac and the diffuse runtime execution engine in Jaseci to build sophisticated AI products with significantly reduced complexity and teaming.

\section{The Case for a New Computational Model}\label{sec:motivation}

Though recent advancements in serverless computing has been instrumental in improving the ability of teams to more rapidly develop software, significant challenges remain in the development of cutting edge applications and products in our current compute landscape.
An demonstrative problem domain with this challenge are those characterized by applications that include sophisticated AI pipelines on their critical path.

\begin{figure}[tb]
    \centering
    \includegraphics[width=\linewidth]{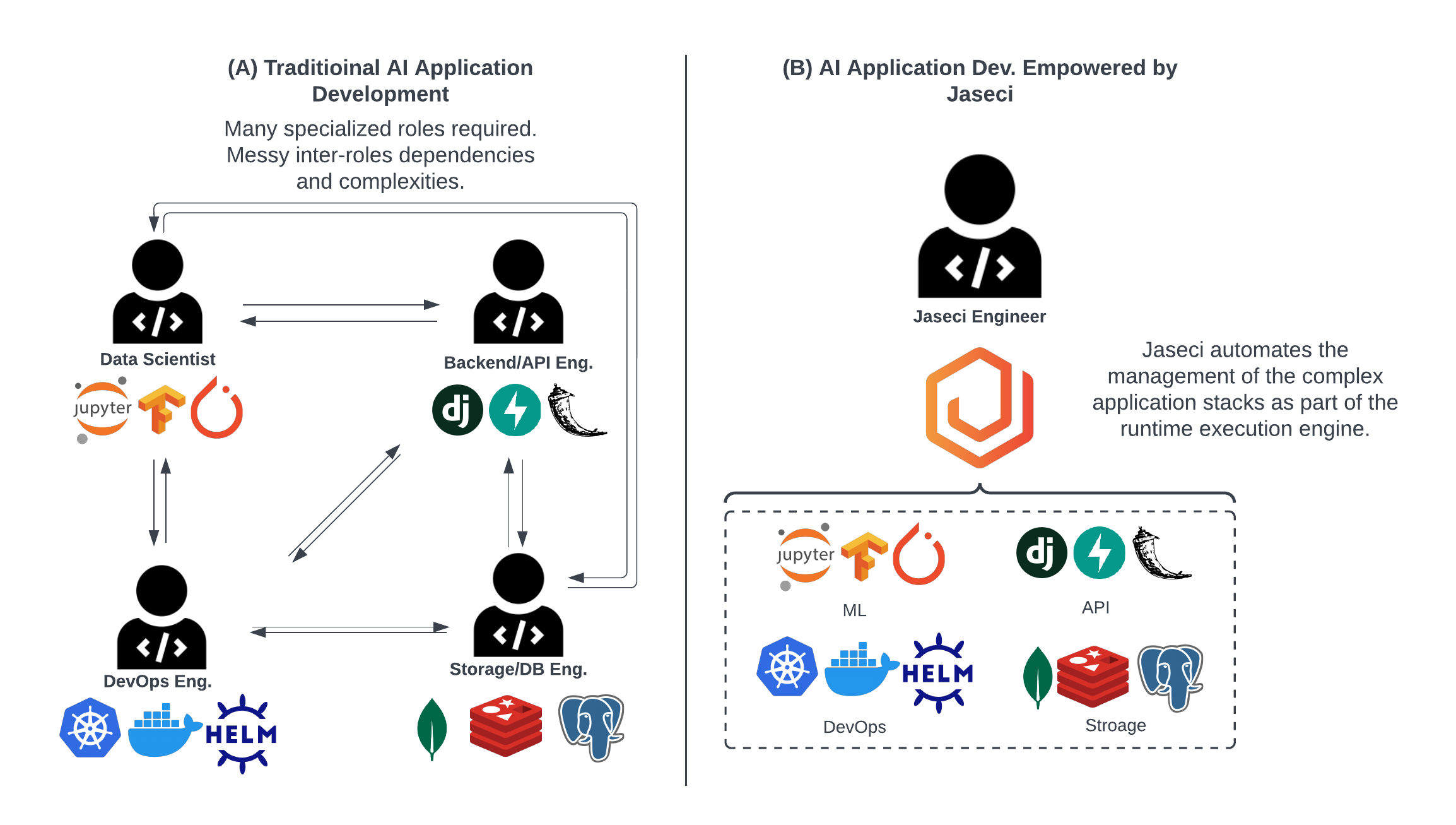}
    \caption{Comparison of typical development team required to realize production grade AI application today (left), and the ability of a single software developer to realize such an application with Jaseci (right). }
    \label{fig:dev}
\end{figure}

\subsection{Problem Scenario}
Figure~\ref{fig:dev}A shows the typical set of  often siloed roles needed to create software in this environment.
The first critical role needed is an \emph{architect / tech-lead} responsible for architecting the software solution across disparate components, programming languages, frameworks, and SDKs.
If a microservice ecosystem is needed (which is a must for modern
AI applications), the architect will also decide what will and won't be its own service (container) and define the interfaces between these disparate  services.
For the AI model work, the role of a \emph{data scientist / ML engineer} is needed.
This role typically works primarily in Jupyter notebooks selecting, creating, training and tuning ML models to support application features.
Production software engineering is typically outside of the scope of this expertise in practice.
The role of a \emph{backend engineer} is needed for implementing the main services of the application and taking the code out of Jupyter notebooks to build the models into the backend (server-side) of the application.
The \emph{backend engineer} is also responsible for supporting new features and creating their API interfaces for \emph{frontend} engineers.
One of the key roles any software team needs to deploy an AI product is a \emph{DevOps engineer}.
This role is solely responsible for deploying and configuring containers to run on a cloud and ensure these containers are operational and scaled to the load requirements of the software.
This responsibility covers configuring software instance pods, database pods, caching layers, logging services, and parameterizing replicas and auto-scaling heuristics.


In this traditional model of software engineering, many challenges and complexity emerge.
An example is the (quite typical) scenario of the first main server-side implementation of the application being a monoservice while DB, caching, and logging are microservices.
As the \emph{ML engineer} introduces models of increasing size, the \emph{dev-ops} person alerts the team that the cloud instances, though designated as \emph{large}, only have 8gb of ram.
Meanwhile new AI models being integrated exceed this limit.
This event leads to a re-architecture of the main monoservice to be split out AI models into microservices and interfaces being designed or adopted leading to significant backend work / delays.
In this work, we aim to create a solution that would move all of this decisioning and work under the purview of the automated runtime system.

Ultimately, the mission of Jaseci is to accelerate and democratize the development and deployments of end-to-end scalable AI applications as presented in Figure~\ref{fig:dev}B.
To this end, we present a novel set of higher level abstractions for programming sophisticated software in a micro-service/serverless AI and a full stack architecture and programming model that abstracts away and automates much of the complexity of building diffuse applications on a distributed compute substrate of potentially thousands of compute nodes.

\section{Data Spacial Programming Model with Jac}\label{sec:jac}
\begin{figure}[tb]
    \centering
    \includegraphics[width=\linewidth]{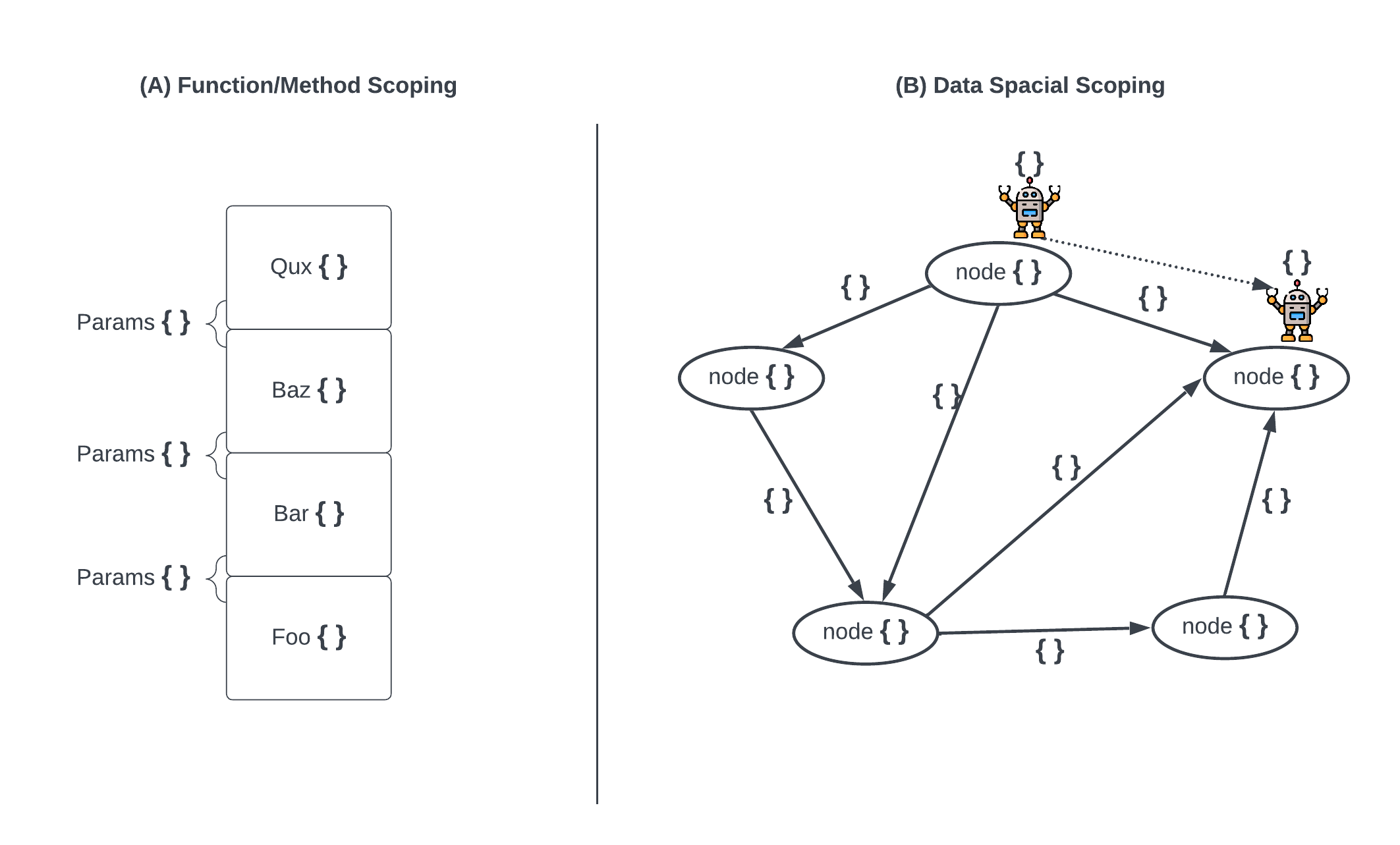}
    \caption{A visualization of the behavior of scopes and problem solving abstractions provided by the near ubiquitous function / method based languages (left) and the data spacial programming model (right).}
    \label{fig:benefit}
\end{figure}

Traditionally in computer science, the task of raising the level abstraction in a computational model has primarily been for the goal of increasing programmer productivity.
This productivity comes from allowing engineers to function at the problem level while hiding the complexity of the underlying system.
The Jac language introduces a set of new abstractions guided by these principles based on two key insights.
First, Jac recognizes the emerging need for programmers to reason about and solve problems with graph representations of data.
Second, Jac further supports the need for algorithmic modularity and encapsulation to change and prototype production software in place of prior running codebases.
Based on these insights, we introduce two new sets of abstractions.
As shown in Figure~\ref{fig:benefit}b, Jac's \textbf{data-spacial scoping} natively facilitates graph based problem solving by replacing the traditional \emph{temporal} notion of scope with a function's activation record with scoping that is flattened and spatially laid out in graph structure.
This type of scoping allows for richer semantics for the organization of the data relevant to the problem being solved.
Figure~\ref{fig:benefit}b also depicts Jac's \textbf{agent oriented programming} as little robots.
Each robot carries scope with it as it walks and performs compute relevant to where it sits on the graph.
These `agent' abstractions capture the need for algorithmic modality and encapsulation when introducing solutions to already sophisticated codebases.
Jac can be used solely to build out complete solutions or as glue code with components built in other languages.
By leveraging these new language abstractions, HomeLendingPal~\cite{hlp-website} was able to create a production grade conversational AI experience with $\sim$300 lines of code in contrast to the tens of thousands it would take to build in a traditional programming language.

\section{Diffuse Runtime Execution with Jaseci}\label{sec:arch}
\begin{figure}[tb]
    \centering
    \includegraphics[width=\linewidth]{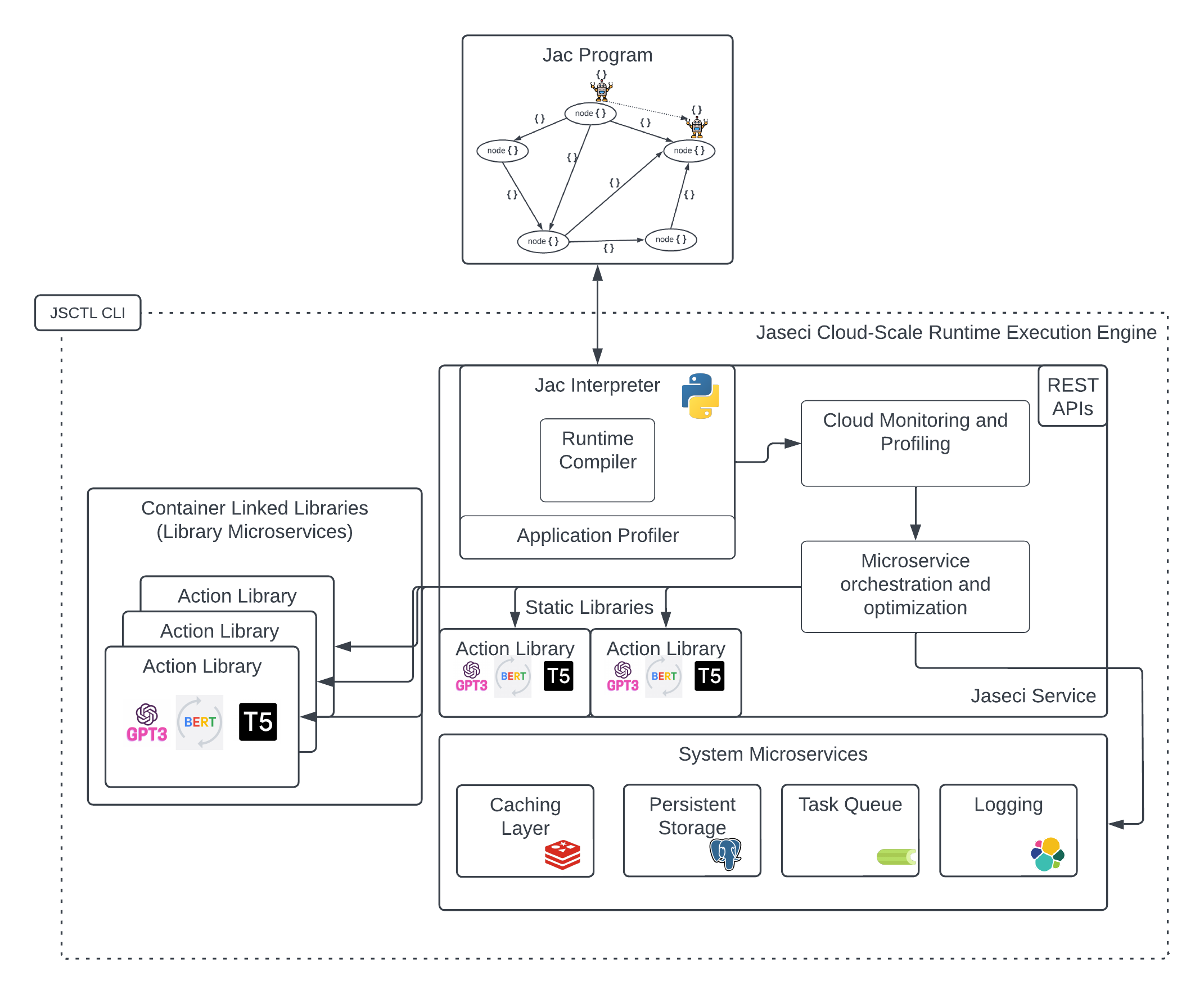}
    \caption{The architecture of the Jaseci diffuse runtime execution. The runtime stack includes and combines information from interpreter level profiling, cloud monitoring and profiling, microservice orchestrator and optimizer. Container linked libraries are also depicted. }
    \label{fig:benefit2}
\end{figure}

Jaseci's cloud-scale runtime engine presents a higher level abstraction of the software stack.
The \emph{diffuse} runtime engine subsumes responsibility not only for the optimization of program code, but also the orchestration, configuration, and optimization of constituent micro services and the full cloud compute stack.
Duties such as container formation, microservice scaling, scheduling and optimization are automated by the runtime.
For example, as shown in Figure~\ref{fig:benefit2}c Jaseci introduces the concept of \textbf{container linked libraries} to complement traditional notions of statically and dynamically linked libraries.
From the programmers perspective, they need not know whether a call to a library is fused with the running programming instance or a remote call to a microservice somewhere in a cluster.
The decisioning of what \emph{should} be a microservice and what should be statically in the programs object scope is made automatically and seamlessly by the Jaseci \textbf{microservice orchestration engine}.
Underlying in-cluster microservices are encapsulated and hidden with this abstraction.
With the runtime having full visibility and control over the diffuse application, high complexity runtime decisions and heuristics such as autoscaling is brought under the purview of the runtime software stack, relieving the need of manual configuration.
With this Jaseci runtime, a single frontend engineer was able to implement the full ZeroShotBot~\cite{zsb-website} application (which uses a number of transformer neural networks) without writing a single line of traditional `backend' code.
This implementation currently support tens of thousands of queries a day across about $\sim$12 business customers with tens of thousands of individual end users in a single production environment.

\section{Case Studies and Discussion}\label{sec:casestudy}

Jaseci is available on Github~\cite{jaseci-github} under MIT open source license and is composed of an ecosystem of tools spanning 3 packages.
These include \textbf{Jaseci Core}, its core execution engine,  \textbf{Jaseci Serv}, its diffuse runtime cloud-scale execution engine, and \textbf{Jaseci Kit}, a collection of cutting edge AI engines provided by the Jaseci community.
In addition to these main codebases, an experimental toolkit we call \textbf{Jaseci Studio} is in development to provide visual programming and debugging tooling for developers building with Jaseci.

There are a number of notable examples of Jaseci's use in production.
These users include  four selected start-up companies that have adopted Jac and Jaseci as their development engine and have already launched their products built using Jaseci.

\indent \textbf{ myca.ai}~\cite{myca-website} - a B2C personal productivity platform that uses AI to understand personal behavior trends and help users allocate their time, prioritize their tasks and achieve personal growth goals. Using Jaseci, myca.ai’s back-end development only took 1 month and myca.ai was launched within 3 months’ development to the public. Myca.ai is one of the fast growing personal growth tool and has received positive feedback from their users.

\indent \textbf{ZeroShotBot}~\cite{zsb-website} - a B2B company that develops a cutting edge conversational AI platform using Jaseci. The product development took 2 months and was done by frontend engineers. Zeroshotbot has gained significant market traction and has been in business discussions with major logos such as Volaris, Pizzahut to provide readily deployable FAQ chatbots.

\indent \textbf{Truselph}~\cite{ts-website} - A minority founded startup. Truselph creates an avatar of the person and builds conversational intelligence that allows the general public to interact with the avatar and ask questions, while the avatar will be able to provide personalized answers with emotions and facial expressions. Truselph is in partnership with Lenovo to co-develop Truselph powered Kiosks for retail stores and is in business discussions with chains such as Sephora.

\indent \textbf{ Home Lending Pal}~\cite{hlp-website} - an AI Powered Mortgage Advisor. Home Lending Pal is a minority founded start-up that helps people, especially under-served minority population to navigate through the mortgage and home purchase process. Home Lending Pal adopted Jaseci to provide two main product features: 1 - personalized mortgage advice and 2 - Kev, an AI-powered chatbot that will answer users questions about the process and give them a plan to improve their finances.

\balance
\section{Conclusion}
Jaseci is a novel computational model invented, designed and implemented to address this challenge.
Jaseci includes a novel programming model we call \emph{data-spacial programming} and a runtime engine we call the \emph{diffuse execution environment} to enable rapid development of large scale and nimble AI applications.
Our initial infrastructure has been used in practice to achieve 10x reduction in development time and near 100\% elimination of typical backend code needed for a complicated AI based application.
Jaseci~\cite{jaseci-website} was open sourced in 2021~\cite{jaseci-github}~\cite{jaseci-pypi}.
Today Jaseci is in production with 4 distinct commercial products built on the engine, including Myca~\cite{myca-website}, HomeLendingPal~\cite{hlp-website},  ZeroShotBot~\cite{zsb-website} and TrueSelph~\cite{ts-website}.

\ifCLASSOPTIONcaptionsoff
    \newpage
\fi

\bibliography{biblio}
\bibliographystyle{IEEEtran}

\end{document}